\documentclass[aps,preprintnumbers,eqsecnum,amsmath,amssymb,showpacs,nofootinbib]{revtex4}
\usepackage{graphicx}\usepackage{dcolumn}\usepackage{bm}\usepackage{epsfig}

\begin{document}
\title{Accuracy of the pion elastic form factor extracted from a local-duality sum rule}
\author{Irina Balakireva$^{1}$, Wolfgang Lucha$^{2}$, Dmitri Melikhov$^{1,2,3}$}
\affiliation{$^1$SINP, Moscow State University, 119991, Moscow, Russia\\
$^2$HEPHY, Austrian Academy of Sciences, Nikolsdorfergasse 18, A-1050, Vienna, Austria\\
$^3$Faculty of Physics, University of Vienna, Boltzmanngasse 5, A-1090, Vienna, Austria}
\date{\today}
\begin{abstract}
We analyze the accuracy of the pion elastic form factor predicted by a local-duality 
(LD)~version~of dispersive sum
rules. To probe the precision of this theoretical approach, we
adopt potential~models with interactions that involve both Coulomb
and confining terms. In this case, the exact form~factor may be
obtained from the solution of the Schr\"odinger equation and
confronted with the LD sum-rule results. We use parameter values
appropriate for hadron physics and observe that, independently~of
the details of the confining interaction, the deviation of the LD
form factor from the exact form~factor culminates in the region
$Q^2\approx4$--$6$ GeV$^2$. For larger $Q^2$, the accuracy of the
LD description~increases rather fast with $Q^2$. A similar picture
is expected for QCD. For the pion form factor, existing~data
suggest that the LD limit may be reached already at the relatively
low values $Q^2=4$--$10$ GeV$^2$.~~Thus, large deviations of the
pion form factor from the behaviour predicted by LD QCD sum rules
for higher values of $Q^2,$ as found by some recent analyses,
appear to us quite improbable. New accurate~data on the pion form
factor at $Q^2=4$--$10$ GeV$^2$ expected soon from JLab will have
important implications for the behaviour of the pion form factor
in a broad $Q^2$ range up to asymptotically large values~of~$Q^2$.
\end{abstract}
\maketitle

\section{\label{Sec:1}Introduction}
\begin{figure}[h!]
\begin{center}
\includegraphics[scale=.75]{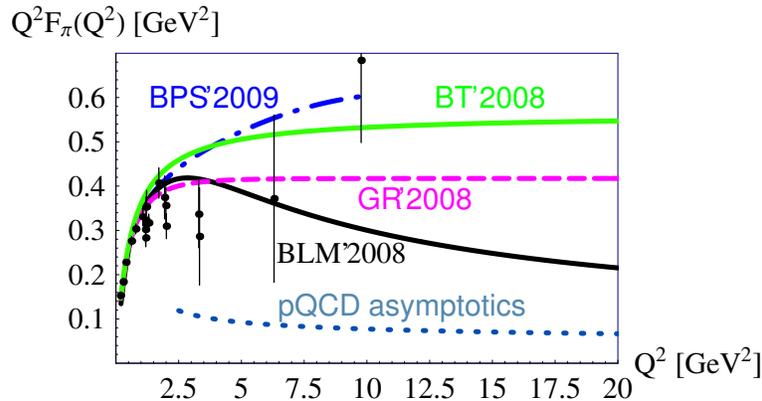}
\caption{\label{Plot:1a}
Some predictions for the pion elastic form
factor $F_\pi(Q^2)$ --- lower solid (black) line:\ BLM'2008
\cite{braguta}, upper solid (green) line:\ BT'2008 \cite{recent1},
dashed (magenta) line:\ GR'2008 \cite{recent2}, and dash-dotted
(blue) line:~BPS'2009 \cite[Eq.~(4.11b)]{bakulev} ---
vs.~experiment~\cite{data_largeQ2,data_JLab}.}
\end{center}
\end{figure}

In spite of a rather long
history of theoretical studies of the pion elastic form factor, no
consensus on its behaviour~in the spacelike region $Q^2\ge2$--$4$
GeV$^2$ has been reached so far. For instance, recent theoretical
investigations \cite{recent1,recent2,recent3}~report results for
the pion form factor much larger than earlier ones
(cf.~Fig.~\ref{Plot:1a}): According to
\cite{recent1,recent2,recent3}, even at
$Q^2\approx50$--$100$~GeV$^2$ the pion form factor $F_\pi(Q^2)$
remains much larger than the asymptotic behaviour expected from
perturbative QCD~\cite{pqcd}
\begin{eqnarray}
\label{ffass}Q^2F_\pi(Q^2)=8\pi\alpha_s(Q^2)f_\pi^2,
\end{eqnarray}
with $f_\pi$ the decay constant of the pion and $\alpha_s(Q^2)$
the strong coupling. Subleading logarithmic and power corrections
modify the behaviour (\ref{ffass}) at large but finite~$Q^2$. In
early applications~of QCD one hoped that power corrections~would
vanish fast enough with $Q^2$; however, later investigations
revealed that nonperturbative power corrections dominate~the form
factor $F_\pi(Q^2)$ up to relatively high $Q^2\approx10$--$20$
GeV$^2$. This picture has arisen from different approaches
\cite{isgur,simula,m,anisovich,roberts,braun}. At this stage, the
conclusion was that even for $Q^2$ as large as $Q^2=20$ GeV$^2$
the $O(1)$ term provides about half of~the form factor and the
perturbative-QCD formula based on factorization starts to work
well only at $Q^2\ge50$--$100$~GeV$^2$.

A convincing and largely model-independent argument comes from QCD
sum rules in their local-duality (LD)~version \cite{nesterenko},
which enables one to take into account, on an equal footing, the
$O(1)$ and the $O(\alpha_s)$ contributions to $F_\pi(Q^2)$: the LD
form of QCD sum rules predicts in an essentially model-independent
way the relative weights of these contributions \cite{braguta},
but needs an additional input --- the $Q^2$-dependent effective
threshold $s_{\rm eff}(Q^2)$ --- in order to calculate
$F_\pi(Q^2)$. All previous applications of LD sum rules had to
rely on assumptions about the $Q^2$ behaviour of the effective
threshold. The LD {\it model\/} \cite{nesterenko,ld} assumes that
reasonable estimates for the form factor already starting from
$Q^2\ge2$--$4$ GeV$^2$~may be obtained by setting $s_{\rm
eff}(Q^2)$ equal to its LD value $s_{\rm LD}=4\pi^2f_\pi^2$. Under
this assumption, $F_\pi(Q^2)$ has been calculated
in~\cite{braguta}.\footnote{Notice that the pioneering work on the
LD model \cite{nesterenko} used a simple interpolation formula for
the unknown $O(\alpha_s)$ contribution to~$F_\pi(Q^2).$ The first
analysis of the pion form factor taking into account the
$O(\alpha_s)$ contribution obtained in \cite{braguta1} was
performed in \cite{braguta}.}

\begin{figure}[hb!]\begin{center}
\includegraphics[scale=.75]{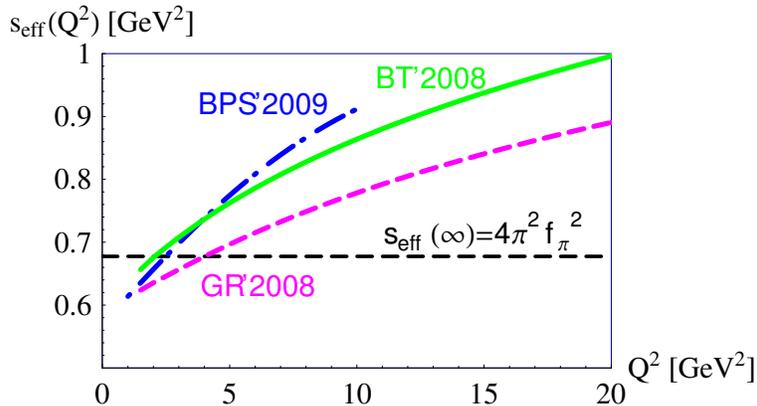}
\caption{\label{Plot:1b}Equivalent effective thresholds $s_{\rm
eff}(Q^2)$ resulting from the findings of
\cite{recent1,recent2,bakulev}, identified by the same line codes
as~in~Fig.~\ref{Plot:1a}.}\end{center}\end{figure}

Surprisingly, recent studies
\cite{recent1,recent2,recent3,bakulev} find large deviations from
the LD results \cite{braguta}. To quantify these
deviations,~Fig.~\ref{Plot:1b} compares the equivalent effective
thresholds\footnote{The {\em equivalent effective threshold\/} is
defined be requiring that a given theoretical prediction for the
pion form factor is reproduced~by~the LD sum rule (\ref{SR_G5}) if
the corresponding equivalent threshold is used (cf.~Sect.~2).}
computed from the results of these analyses. For some of these
studies, the~deviation of the equivalent threshold from $s_{\rm
LD}$ rises with $Q^2$, although the accuracy of the LD model is
expected to increase~with~$Q^2$. However, all of these studies
involve explicit or implicit assumptions.

The goal of this paper is to test the accuracy of the LD model by
exploiting the fact that in quantum mechanics the form factor may
be found in two ways, {\em viz.}, by the LD sum rule and by
solving the Schr\"odinger equation. For the first time we
calculate the exact $Q^2$-dependent threshold in a
quantum-mechanical potential model for different~confining
interactions. We find that the accuracy of the LD approximation at
a given value of $Q^2$ depends on the details of the confining
interaction. More importantly, we observe a universal feature that
does not depend on these details:~namely, for realistic values of
the parameters of the potential model relevant for hadron physics,
the accuracy of the LD model increases with $Q^2$ already starting
at relatively low values $Q^2\approx4$--$6$ GeV$^2$.

In QCD, experimental data on the pion form factor at small $Q^2$
indicate that the LD limit may be reached already at relatively
low values of $Q^2\approx5$--$6$ GeV$^2.$ This, of course, does
not mean that the asymptotic pQCD formula describes well the form
factor in this region: the $O(1)$ term still dominates the form
factor up to $Q^2\approx20$ GeV$^2.$ Therefore, we conclude that
large and growing-with-$Q^2$ deviations from the LD limit in the
region $Q^2\approx20$--$50$ GeV$^2$ implied by the recent analyses
\cite{recent1,recent2,recent3} seem to us very unlikely.

This paper is organized as follows: In the next section, after
recalling rather well-known basics of the LD~limit~of the QCD
sum-rule approach we formulate, for use in these LD sum rules, a
model for $s_{\rm eff}(Q^2)$ slightly more sophisticated than the
one used in \cite{braguta}. In Sec.~\ref{Sec:3}, we discuss in
detail rigorous features and assumptions employed in the
LD~approach. Section \ref{Sec:4} presents the quantum-mechanical
counterpart of the LD model for $F_\pi(Q^2)$ using potentials that
involve both Coulombic and confining interactions: we examine the
impact of the specific shape of the confining interactions on the
accuracy of the LD model for the elastic form factor. Section
\ref{Sec:5} summarizes our conclusions and
outlook.~~Appendix~\ref{Sec:App} presents several technical
details of perturbative two-loop calculations in quantum
mechanics.

\section{\label{Sec:2}Sum rule}
The fundamental objects for a
sum-rule extraction of pion features are the two- and three-point
correlation functions\begin{align}\label{sr_pion}
\Pi\!\left(p^2\right)&=\int{\rm d}^4x\,e^{{\rm i}px}
\left<\Omega\left|T[j(x)\,j^\dag(0)]\right|\Omega\right>,
\nonumber\\ \Gamma\!\left(p^2_1,p^2_2,q^2\right)&=\int{\rm
d}^4x_1\,{\rm d}^4x_2\,e^{{\rm i}(p_1x_1-p_2x_2)}
\left<\Omega\left|T[j(x_1)\,J(0)\,j^\dag(x_2)]\right|\Omega\right>,
\qquad q\equiv p_1-p_2,\qquad Q^2\equiv-q^2;\end{align}where
$\Omega$ labels the physical vacuum, $j$ is shorthand for the
interpolating axial current $j_{5\alpha}$ of the positively
charged pion,
$\left<\Omega\left|j_{5\alpha}(0)\right|\pi(p)\right>={\rm
i}p_{\alpha}f_{\pi}$, $J$ labels the electromagnetic current
$J_{\mu}$, and for brevity we omit all Lorentz indices. In QCD,
the correlators (\ref{sr_pion}) can be found by applying their
OPEs. Instead of discussing the Green
functions~(\ref{sr_pion})~in~Minkowski space, it is convenient to
study the time-evolution operators in Euclidean space, which arise
upon performing the Borel transformation $p^2\to\tau$ to a
parameter $\tau$ related to Euclidean time. The Borel image of the
two-point correlator~$\Pi(p^2)$~is\begin{eqnarray}\label{SR_P2}
\Pi_{\rm OPE}(\tau)=\int\limits_0^\infty{\rm
d}s\,e^{-s\tau}\,\rho_{\rm pert}(s)+\Pi_{\rm
cond}(\tau),\qquad\rho_{\rm pert}(s)=
\rho_0(s)+\alpha_s\,\rho_1(s)+O(\alpha^2_s),\end{eqnarray}with
spectral densities $\rho_i(s)$ related to perturbative two-point
graphs, and nonperturbative power corrections $\Pi_{\rm
cond}(\tau)$. At hadron level, insertion of intermediate hadron
states casts the Borel-transformed two-point correlator into the
form\begin{equation}\label{SR_P3}\Pi(\tau)
=f_\pi^2\,e^{-m_\pi^2\tau}+\mbox{excited states}.\end{equation}In
this expression for $\Pi(\tau),$ the first term on the right-hand
side constitutes the pion contribution. Applying the~double Borel
transform $p^2_{1,2}\to\tau/2$ to the three-point correlator
$\Gamma(p^2_1,p^2_2,q^2)$ results, at QCD level,~in
\begin{align}
\label{3PF}
\Gamma_{\rm OPE}(\tau,Q^2)&=
\int\limits_0^\infty\,\int\limits_0^\infty{\rm d}s_1\,{\rm d}s_2
\exp\!\left(-\frac{s_1+s_2}{2}\,\tau\right)\Delta_{\rm
pert}(s_1,s_2,Q^2)+\Gamma_{\rm cond}(\tau,Q^2),\nonumber\\
\Delta_{\rm pert}(s_1,s_2,Q^2)&=\Delta_0(s_1,s_2,Q^2)
+\alpha_s\,\Delta_1(s_1,s_2,Q^2)+O(\alpha^2_s),
\end{align}
where
$\Delta_{\rm pert}(s_1,s_2,Q^2)$ is the double spectral density of
the three-point graphs of perturbation theory and $\Gamma_{\rm
cond}(\tau,Q^2)$ labels the power corrections. Inserting
intermediate hadron states yields, for the hadron-level
expression~for~$\Gamma(\tau,Q^2),$
\begin{equation}
\Gamma(\tau,Q^2)=F_\pi(Q^2)\,f_\pi^2\,e^{-m_\pi^2\tau}
+\mbox{excited states}.
\end{equation}

{\it Quark--hadron duality assumes\/} that above effective
continuum thresholds $s_{\rm eff}$ the excited-state contributions
are~dual to the high-energy regions of the perturbative graphs. In
this case, the relevant sum rules read in the chiral
limit~\cite{nesterenko,ioffe}
\begin{align}
\label{SR_P4}
f_\pi^2&=\int\limits_0^{\bar s_{\rm eff}(\tau)}{\rm d}s\,
e^{-s\tau}\,\rho_{\rm
pert}(s)+\frac{\left<\alpha_s\,G^2\right>}{12\pi}\,\tau
+\frac{176\pi\,\alpha_s\left<\bar qq\right>^2}{81}\,\tau^2+\cdots,\\ 
\label{SR_G4}
F_\pi(Q^2)\,f_\pi^2&=\int\limits_0^{s_{\rm
eff}(Q^2,\tau)}\,\int\limits_0^{s_{\rm eff}(Q^2,\tau)}{\rm
d}s_1\,{\rm d}s_2\, \Delta_{\rm pert}(s_1,s_2,Q^2)
\exp\!\left(-\frac{s_1+s_2}{2}\,\tau\right)\nonumber\\
&+\frac{\left<\alpha_s\,G^2\right>}{24\pi}\,\tau
+\frac{4\pi\,\alpha_s\left<\bar qq\right>^2}{81}\,\tau^2
\left(13+Q^2\,\tau\right)+\cdots.
\end{align}

As a consequence of the use of local condensates, the right-hand
side of (\ref{SR_G4}) involves polynomials in $Q^2$ and~therefore
increases with $Q^2$, whereas the form factor $F_\pi(Q^2)$ on the
left-hand side should decrease with $Q^2.$ Therefore, at large
$Q^2$ the sum rule (\ref{SR_G4}), with its truncated series of
power corrections, cannot be directly used. There are
essentially~only~two ways for considering the region of large
$Q^2$.

One remedy is the resummation of all power corrections: the
resummed power corrections decrease with increasing $Q^2.$ This
may be achieved by the introduction of nonlocal condensates
\cite{nonlocal} in a, however, model-dependent
manner~\cite{chernyak_2006}.

Another --- rather simple --- option is to fix the Borel parameter
$\tau$ to the value $\tau=0,$ thus arriving at a {\it
local-duality sum~rule\/}~\cite{nesterenko}. Therein all power
corrections vanish and the remaining perturbative term decreases
with $Q^2$. In the~LD limit, one finds\begin{align}\label{SR_P5}
f_\pi^2&=\int\limits_0^{\bar s_{\rm eff}}{\rm d}s\,\rho_{\rm
pert}(s)=\frac{\bar s_{\rm eff}}{4\pi^2}
\left(1+\frac{\alpha_s}{\pi}\right)+O\!\left(\alpha_s^2\right),\\
\label{SR_G5}F_\pi(Q^2)\,f_\pi^2&=\int\limits_0^{s_{\rm
eff}(Q^2)}\,\int\limits_0^{s_{\rm eff}(Q^2)}{\rm d}s_1\,{\rm
d}s_2\, \Delta_{\rm pert}(s_1,s_2,Q^2).\end{align}The spectral
densities $\rho_{\rm pert}(s)$ and $\Delta_{\rm
pert}(s_1,s_2,Q^2)$ are calculable by perturbation theory. Hence,
by fixing $\bar s_{\rm eff}$ and $s_{\rm eff}(Q^2)$, it is
straightforward to extract the pion's decay constant $f_\pi$ and
form factor $F_\pi(Q^2).$

Noteworthy, the effective and the physical thresholds are
different quantities: The latter is a constant determined by the
masses~of the hadron states. The effective thresholds $\bar s_{\rm
eff}$ and $s_{\rm eff}$ are parameters of the sum-rule
method~related~to the specific realization of quark--hadron
duality; in general, they are {\em not\/} constant but depend on
external kinematical variables \cite{lms_sr1,lms_sr2}.

Let us recall the important properties of the spectral densities
on the right-hand sides of (\ref{SR_P2}) and (\ref{3PF}):
For~$Q^2\to0,$ the Ward identity relates the spectral densities
$\rho_i(s)$ and $\Delta_i(s_1,s_2,Q^2)$ of two- and three-point
functions~to~each~other:
\begin{eqnarray}
\label{Ward_identity}
\lim_{Q^2\to0}\Delta_i(s_1,s_2,Q^2)=\rho_i(s_1)\,\delta(s_1-s_2),\qquad
i=0,1,\dots.
\end{eqnarray}
For $Q^2\to\infty$ and $s_{1,2}$ kept
fixed, explicit calculations \cite{braguta1} yield\begin{eqnarray}
\label{large_momentum}\lim_{Q^2\to\infty}\Delta_0(s_1,s_2,Q^2)\propto
\frac{1}{Q^4},\qquad\lim_{Q^2\to\infty}\Delta_1(s_1,s_2,Q^2)
=\frac{8\pi}{Q^2}\,\rho_0(s_1)\,\rho_0(s_2).
\end{eqnarray}

For the pion form factor $F_\pi(Q^2)$ on the left-hand side of
(\ref{SR_G4}) two exact features are known, namely, its
normalization condition related to current conservation, requiring
$F_\pi(0)=1,$ and the factorization theorem (\ref{ffass}).
Obviously, if we~set
\begin{equation}
\label{seff-sL} 
s_{\rm eff}(Q^2\to0)=\frac{4\pi^2\,f^2_{\pi}}{1+\alpha_s(0)/\pi},\qquad
s_{\rm eff}(Q^2\to\infty)=s_{\rm LD}\equiv4\pi^2\,f_\pi^2,
\end{equation}
the form factor $F_\pi(Q^2)$ extracted from the LD
sum rule (\ref{SR_G5}) satisfies both of these rigorous
constraints. At small~$Q^2,$ we assume a freezing of
$\alpha_s(Q^2)$ at the level 0.3, as is frequently done.

\begin{figure}[ht!]
\begin{center}
\includegraphics[scale=.75]{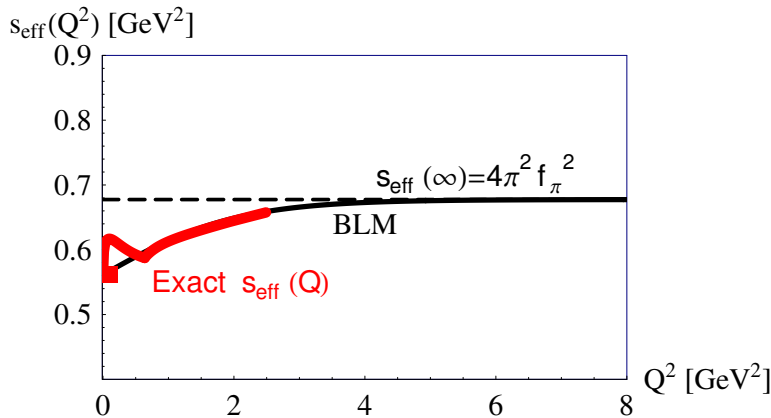}
\caption{\label{Plot:1c}
Effective continuum threshold $s_{\rm eff}(Q^2)$ as function of $Q^2$. 
Bold (red) line: exact behaviour
$s_{\rm eff}(Q^2)$ as reconstructed from the experimental data for
$F_\pi(Q^2)$ \cite{data_largeQ2,data_JLab} for $Q^2\le2.5$
GeV$^2$. Solid (black) line: our simple interpolating model
(\ref{seff_LD})~for~$s_{\rm eff}(Q^2)$.}
\end{center}
\end{figure}

With these preliminaries at our disposal, we are now in the
position to formulate our approach to the pion elastic~form factor
$F_\pi(Q^2)$ within the framework of QCD sum rules in the LD limit:
\begin{enumerate}
\item 
We extract $F_\pi(Q^2)$ from the
dispersive three-point QCD sum rule in LD limit (\ref{SR_G4}): the
latter has the favourable feature that all power corrections
vanish and all details of nonperturbative dynamics are encoded in
one quantity, the effective threshold $s_{\rm eff}(Q^2)$. We take
into account the perturbative spectral densities up to
$O(\alpha_s)$ accuracy.
\item It is easy to construct a model for
$s_{\rm eff}(Q^2)$ by some smooth interpolation between its values
at $Q^2=0,$ defined by the Ward identity, and $Q^2\to\infty$,
determined by factorization: a simple parametrization with a
single~constant~$Q_0$ fixed by fitting the data at $Q^2=1$ GeV$^2$
might~read
\begin{equation}
\label{seff_LD}s_{\rm
eff}(Q^2)=\frac{4\pi^2\,f^2_\pi}{1+\alpha_s(0)/\pi}\left[1
+\tanh\!\left(\frac{Q^2}{Q_0^2}\right)\frac{\alpha_s(0)}{\pi}\right],
\qquad Q_0^2=2.02\;{\rm GeV}^2.
\end{equation}
According to
Fig.~\ref{Plot:1c}, our interpolation perfectly describes the
well-measured data in the range $Q^2\approx0.5$--$2.5$~GeV$^2.$
\end{enumerate}Note that the effective continuum threshold $s_{\rm
eff}(Q^2)$ in Eq.~(\ref{seff_LD}) approaches its limit $s_{\rm
LD}$ already at $Q^2\approx4$--$5$ GeV$^2$. For $Q^2>4$--$5$
GeV$^2$, it practically coincides with the LD effective threshold
of \cite{nesterenko}. Moreover, for $Q^2>5$--$6$ GeV$^2$~the
formula~(\ref{seff_LD}) is pretty close to the model of
\cite{braguta}.\footnote{It goes without saying that our present
goal is {\em not\/} to improve the model of \cite{braguta} but to
probe the accuracy of this model as a function~of~$Q^2$.} Hence,
also the resulting prediction for $F_\pi(Q^2)$ is rather close
to~the one~we~found earlier \cite{braguta}
(cf.~Fig.~\ref{Plot:1a}). Obviously, the model labeled BLM in
Fig.~\ref{Plot:1a} provides a perfect description of the available
$F_\pi(Q^2)$ data in the region $Q^2=1$--$2.5$ GeV$^2$. For
$Q^2\ge3$--$4$ GeV$^2$, it reproduces well all the
data,~except~for~a point at $Q^2=10$ GeV$^2,$ where it is off the
present experimental value, which anyhow has a rather large error,
by~some two standard deviations.\footnote{It is virtually
inconceivable to construct models compatible with all experimental
results within $Q^2=2.5$--$10$ GeV$^2,$ as revealed by closer
inspection of Fig.~\ref{Plot:1a}: those approaches which hit the
data at $Q^2=10$ GeV$^2$ overestimate the better-quality data
points~at~$Q^2\approx2$--$4$~GeV$^2$.}

\section{\label{Sec:3}What is predicted by and what is conjectured in the LD model?}
Interestingly, in the region
$Q^2\ge3$--$4$ GeV$^2$ the BLM model yields considerably lower
predictions than the results of the different theoretical
approaches presented in Refs.~\cite{recent1,recent2,recent3,bakulev}. 
For the $F_\pi(Q^2)$
predictions of \cite{recent1,recent2,bakulev}, the~corresponding
equivalent effective thresholds $s_{\rm eff}(Q^2)$ recalculated
from (\ref{SR_G5}) are depicted in Fig.~\ref{Plot:1b}: In all
instances, they considerably exceed, for larger $Q^2,$ the LD
limit $s_{\rm LD}$ dictated by factorization. Moreover, their
deviation from $s_{\rm LD}$ increases with~$Q^2$.

Let us inspect more carefully what is, in fact, predicted by the
LD sum rule and what is conjectured in this approach.

\begin{figure}[ht!]
\begin{center}
\includegraphics[scale=.75]{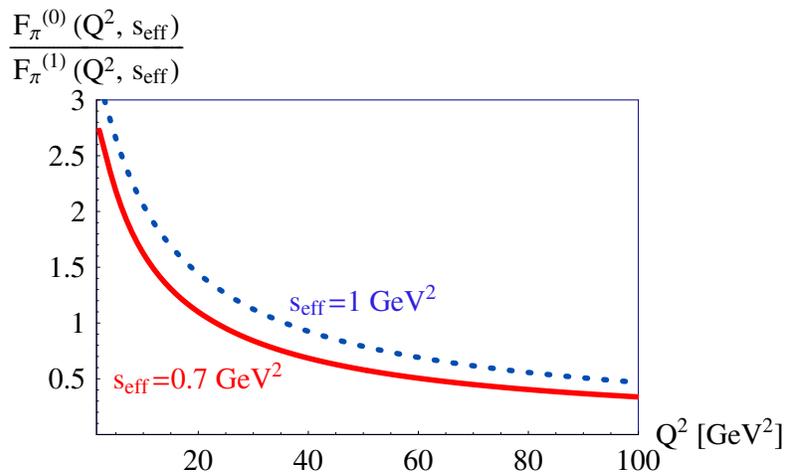}
\caption{\label{Plot:1d}
Ratio $F_\pi^{(0)}(Q^2)/F_\pi^{(1)}(Q^2)$
of $O(1)$ and $O(\alpha_s)$ contributions to the pion elastic form
factor vs.~$Q^2$ for two reasonable~values of the --- here by
assumption constant --- effective threshold $s_{\rm eff}$.}
\end{center}
\end{figure}

The sum rule (\ref{SR_G5}) for the pion form factor relies on two
ingredients: first, on the {\it rigorous calculation\/} of the
spectral densities of the perturbative-QCD diagrams (recall that
power corrections vanish in the LD limit $\tau=0$);~second, on~the
{\it assumption\/} of quark--hadron duality, which claims that the
contributions of the hadronic continuum states may be~well
described by the diagrams of perturbation theory above some
effective threshold $s_{\rm eff}$. Thus, the only ---
although~really essential --- ingredient of the LD sum rule for
the pion elastic from factor is this effective continuum threshold
$s_{\rm eff}(Q^2).$ Let us emphasize that, since the $O(1)$ and
$O(\alpha_s)$ contributions to the pion form factor are governed
by one and the same effective threshold $s_{\rm eff}(Q^2)$, the
relative weights of these contributions may be {\it predicted}.
Their~ratio~$F_\pi^{(0)}(Q^2)/F_\pi^{(1)}(Q^2)$ turns out to be
relatively stable with respect to $s_{\rm eff}$ and may therefore
be calculated relatively accurately (see Fig.~\ref{Plot:1d}).

Quark--hadron duality implies that the effective threshold
(although being a function of $Q^2$) always --- that is,~also for
large $Q^2$ --- stays in a region close to $1$ GeV$^2.$ Moreover,
in order to satisfy the QCD factorization theorem for~the
$O(\alpha_s)$ contribution to the form factor, the effective
threshold should behave like $s_{\rm eff}(Q^2)\to4\pi^2f_\pi^2$
for $Q^2\to\infty.$ This requirement has immediate consequences
for the large-$Q^2$ behaviour of different contributions to the
pion form factor:
\begin{enumerate}
\item 
Since $s_{\rm eff}(Q^2)$
is bounded from above, the $O(1)$ contribution $F_\pi^{(0)}(Q^2)$
to the elastic form factor $F_\pi(Q^2)$ of~the~pion behaves like
$F_\pi^{(0)}(Q^2)\propto1/Q^4$ for $Q^2\to\infty.$ We would like
to emphasize that the decrease of the soft contribution to the
pion elastic form factor like $1/Q^4$ is a~direct consequence of
perturbation theory and quark--hadron duality.
\item 
Consequently,
for large $Q^2$ the pion elastic form factor $F_\pi(Q^2)$ is
dominated by the $O(\alpha_s)$ contribution $F_\pi^{(1)}(Q^2).$
\end{enumerate}

Now, in the holographic models of \cite{recent1,recent2} merely
the soft contribution is considered: it behaves like $1/Q^2$ for
large~$Q^2$. This is only possible if the effective threshold
$s_{\rm eff}(Q^2)$ rises with $Q^2.$ However, this immediately
leads to the violation of the factorization theorem for the
$O(\alpha_s)$ contribution, which is governed by the same
effective threshold. Consequently, we would like to emphasize that
the findings of \cite{recent1,recent2} would imply that the QCD
factorization theorem is violated.~~We thus conclude that the
predictions of \cite{recent1,recent2} for the pion form factor
seem to us improbable.\footnote{A way out would be to assume
different effective thresholds for the $O(1)$ and the
$O(\alpha_s)$ contributions to the pion form factor. This~seems,
however, a rather artificial construction.}

Needless to say, the {\em conventional\/} LD model \cite{ld} for
the effective continuum threshold $s_{\rm eff}(Q^2)$, defined by
the~choice $s_{\rm eff}(Q^2)=4\pi^2f_\pi^2$ for all $Q^2,$ or the
slightly more sophisticated approach of \cite{braguta} are
approximations which do not account for all the subtle details of
the confinement dynamics. Consequently, it is of utmost importance
to acquire a satisfactory understanding of the accuracy to be
expected within this approach, in other words, to obtain some
reliable estimate~of the expected deviations of the exact $s_{\rm
eff}(Q^2)$ from its LD limit $s_{\rm LD}$ in the momentum region
$Q^2\ge4$--$6$ GeV$^2$. To this end, let us take advantage of the
fact that in quantum mechanics all the bound-state features may be
found~exactly~by solving the~Schr\"odinger~equation. On the other
hand, also within quantum mechanics we may construct LD sum rules.

\section{\label{Sec:4}Local-duality model for the elastic form factor in quantum mechanics}
Factorization of hard form factors for
large momentum transfers is the main relevant ingredient of the
approach to the pion form factor advocated in \cite{braguta}.
Hence, for any interaction of Coulomb-plus-confining type this
model~can be tested~in quantum mechanics, which has already proven
to be a rather efficient tool for studying various
features~of~QCD~\cite{qmsr,ms_inclusive,orsay,lms_sr1,lms_sr2}.

Within potential models, the elastic form factor of the ground
state, $F(Q),$ is given in terms of the wave function~$\Psi$~by
\begin{equation}
\label{QM_ff}
F(Q)=\int{\rm d}^3r\,\exp({\rm i}\bm{q}\cdot\bm{r})\,
|\Psi(\bm{r})|^2=\int{\rm
d}^3k\Psi(\bm{k})\,\Psi(\bm{k}+\bm{q}),\qquad Q\equiv|\bm{q}|,
\end{equation}
where the bound-state wave function $\Psi$ is
computed by solving the stationary Schr\"odinger equation for the
Hamiltonian
\begin{equation}
\label{QM_H}
H=\frac{\bm{k}^2}{2m}-\frac{\alpha}{r}+V_{\rm conf}(r),\qquad r\equiv|\bm{r}|.
\end{equation}
At large values of $Q$, the
asymptotic behaviour of the elastic form factor $F(Q)$ is given by
the factorization theorem~\cite{brodsky}:
\begin{equation}
\label{QM_factorization}
F(Q)\xrightarrow[Q\to\infty]{}F_\infty(Q)
\equiv\frac{16\pi\,\alpha\,m\,R_g}{Q^4},\qquad
R_g\equiv|\Psi(\bm{r}=\bm{0})|^2.
\end{equation}
The quantum-mechanical LD sum rules for decay constant $R_g$ and form
factor $F(Q)$ are rather similar to those~in~QCD:
\begin{align}
\label{2pt_QM}
R_g&=\int_0^{\bar k_{\rm eff}}{\rm d}k\,\rho^{\rm QM}_{\rm pert}(k),\\
\label{3pt_QM}F_g(Q)\,R_g&=\int_0^{k_{\rm eff}(Q)}\,\int_0^{k_{\rm eff}(Q)}
{\rm d}k_1\,{\rm d}k_2\,\Delta^{\rm QM}_{\rm pert}(k_1,k_2,Q).
\end{align}
The
spectral densities, $\rho^{\rm QM}_{\rm pert}$ and $\Delta^{\rm
QM}_{\rm pert},$ are calculated from two- and three-point diagrams
of nonrelativistic field~theory; the derivation of these
expressions up to $O(\alpha_s)$ accuracy is presented in Appendix A.

The factorization limit (\ref{QM_factorization}) means for the
momentum-dependent effective threshold in the three-point sum
rule~(\ref{3pt_QM})
\begin{equation}
\label{QM_LD}
k_{\rm eff}(Q)\xrightarrow [Q\to\infty]{}k_{\rm LD}\equiv(6\pi^2\,R_g)^{1/3}.
\end{equation}
For intermediate $Q$, the behaviour of $F(Q)$ is
controlled by the details of the confining interaction via the
corresponding wave function $\Psi$. The LD model assumes that,
also for intermediate $Q,$ one may find a reasonable estimate for
the form factor by setting $k_{\rm eff}(Q)=k_{\rm LD}.$ Hence,
similar to QCD the only property of the bound state which
determines~the form factor in the LD framework is $R_g.$ To probe
the accuracy of this approximation, we consider a set of
confining~potentials
\begin{equation}
\label{QM_conf}
V_{\rm conf}(r)=\sigma_n\,(m\,r)^n, \qquad n=2,1,1/2.
\end{equation}
We
adopt parameter values appropriate for hadron physics, i.e.,
$m=0.175$ GeV for the reduced constituent light-quark mass and
$\alpha=0.3$, and adapt the strengths $\sigma_n$ in our confining
interactions such that the Schr\"odinger equation~yields~for each
potential the same $\Psi(\bm{r}=\bm{0})=0.078$ GeV$^{3/2},$ which
holds for $\sigma_2=0.71$ GeV, $\sigma_1=0.96$ GeV, and
$\sigma_{1/2}=1.4$~GeV.

\begin{figure}[h!]
\begin{center}
\begin{tabular}{cc}
\includegraphics[scale=.717]{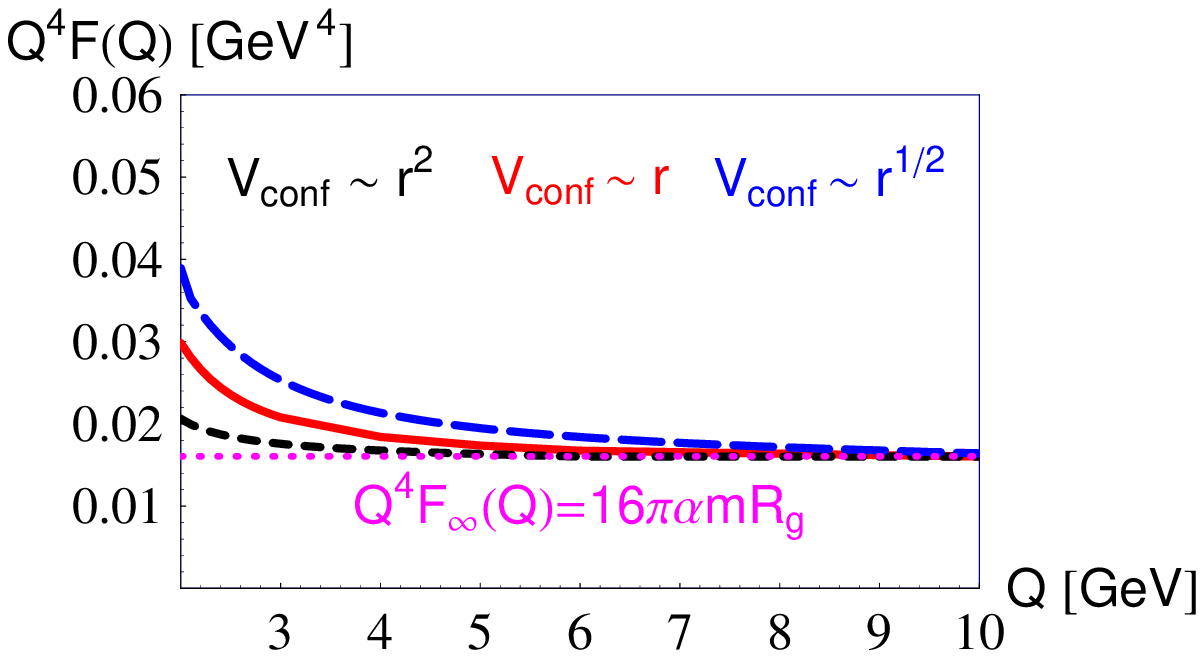}&
\includegraphics[scale=.717]{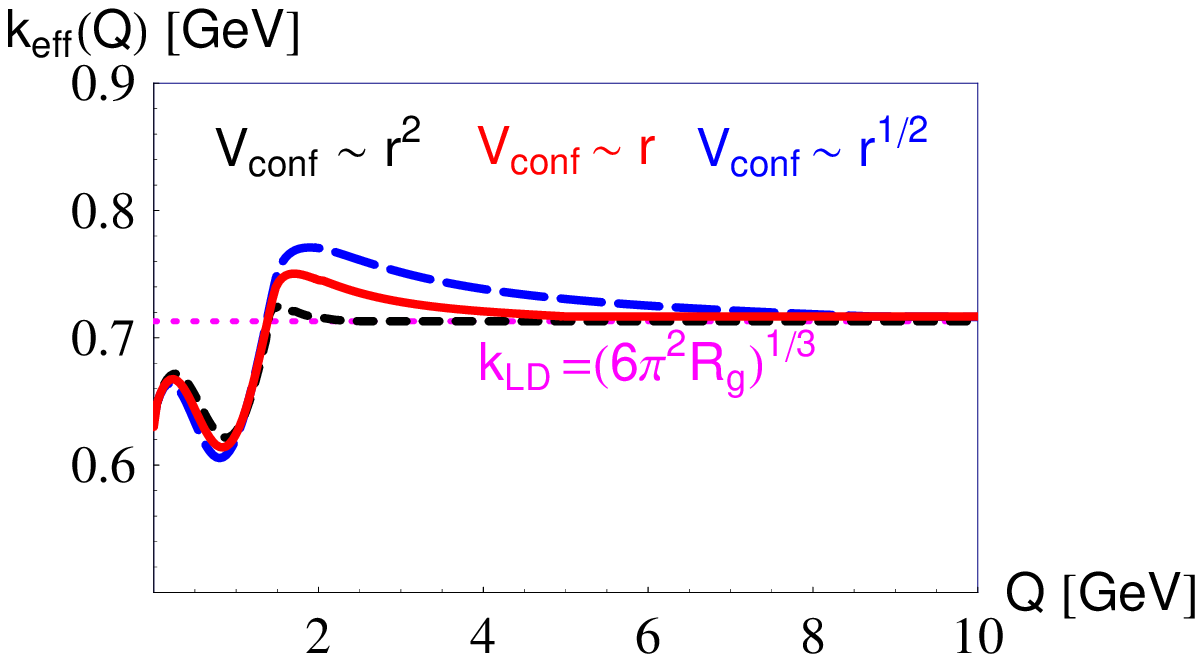}
\\(a)&(b)
\end{tabular}
\caption{\label{Plot:2}(a) Exact vs.\ LD ground-state
form factors $F(Q)$. (b) Exact vs.\ LD effective thresholds
$k_{\rm eff}(Q)$. Dotted (magenta) lines: local-duality limit;
short-dashed (black) lines: harmonic-oscillator confinement,
$n=2;$ full (red) lines: linear confinement,~$n=1;$ long-dashed
(blue) lines: square-root confinement, $n=1/2.$}
\end{center}
\end{figure}

Figure \ref{Plot:2} summarizes our findings in quantum mechanics.
In the region $Q<2$ GeV, the exact $k_{\rm eff}(Q)$ is a
complicated function of $Q$ (Fig.~\ref{Plot:2}b). Moreover, its
behaviour exhibits features similar to the exact QCD threshold,
recalculated~from pion form-factor data (see Fig.~\ref{Plot:1c}).
In the region $Q>2$ GeV, the exact $k_{\rm eff}(Q)$ is smoothly
approaching its LD~limit~$k_{\rm LD}$. For a steeply rising
confining potential ($n=2$), the LD model works with, e.g., 5\%
accuracy already for $Q>2$~GeV. For a slowly rising confining
potential ($n=1/2$), such high accuracy is reached not before
$Q\approx6$ GeV. However, we identify an important universal
feature that does not depend on any details of the confining
interaction: The accuracy~of~the~LD approximation for the
effective threshold, $k_{\rm eff}(Q)\approx k_{\rm LD},$ as well
as the accuracy of the corresponding elastic form factor increase
with $Q$ in the region $Q^2\ge4\;\mbox{GeV}^2.$ Figure
\ref{Plot:2} illustrates this observation for several
confining~potentials with different large-$r$ behaviour. On the
basis of these findings we are forced to conclude that, if at
$Q^2\approx4$--$8$~GeV$^2$~the LD model provides a good
description of the data, the accuracy of this model won't become
worse at larger values~of~$Q^2$.

The crucial question is to which extent these findings can be
carried over to predictions for the pion elastic form~factor by LD
sum rules in QCD. In spite of subtleties related to evident
differences between the spectral densities in relativistic and
nonrelativistic theories \cite{lms_dispersive}, we are convinced
that the pivotal lesson from quantum mechanics also~applies
to~QCD: Above the relatively low value $Q^2\approx4$--$6$~GeV$^2,$
the accuracy of the approximation $s_{\rm eff}(Q^2)\approx s_{\rm
LD}$ will {\em increase\/}~with~$Q^2.$ The results for the form
factor found in \cite{recent1,recent2,recent3,bakulev} clearly
contradict this observation: the
equivalent~effective~threshold~for these models deviates from the
LD threshold $s_{\rm LD}\equiv4\pi^2\,f_\pi^2$ in the broad range
$Q^2\approx4$--$10$ GeV$^2$ and for \cite{recent1,recent2,recent3}
even up to $Q^2\approx100$ GeV$^2;$ more importantly, the
deviations between the equivalent thresholds and the LD
threshold~rise~with~$Q^2.$

\section{\label{Sec:5}Conclusions and Outlook}
We investigated the expected accuracy of the pion elastic form factor obtained by the
LD version of QCD sum~rules. The LD sum rule for the elastic form
factor of some bound state may be constructed in any theory where
hard exclusive amplitudes satisfy the factorization theorem (that
is, in essence, in any theory with interactions behaving
Coulomb-like at small distances and confining at large distances).
In this approach, the form factor is determined by
two~ingredients: the double spectral densities of the diagrams of
perturbation theory and the $Q^2$-dependent effective~threshold
$s_{\rm eff}(Q^2)$. The effective threshold satisfies rigorous
constraints at $Q^2=0$ and for $Q^2\to\infty$ but is unknown for
intermediate~values of $Q^2$. The LD model assumes that
approximating $s_{\rm eff}(Q^2)$ by its value at $Q^2\to\infty$,
$4\pi^2f_\pi^2$, yields reasonable estimates for the form factor
at not too small $Q^2$. We have tested the accuracy of the LD
model in quantum mechanics, where~the exact effective threshold
may be calculated from the exact form factor, obtained by solving
the Schr\"odinger equation.

The new results reported in this work are the following:
\begin{enumerate}
\item 
For $Q^2\le4$ GeV$^2$, the exact effective
threshold exhibits a rapid variation with $Q^2,$ implying that in
this region the accuracy of the LD model depends on the details of
the confining interactions and cannot be predicted~in~advance.
\item 
For $Q^2\ge4$--$6$ GeV$^2$, independently of the details of the
confining interactions, the {\it maximal\/} deviation of~the~exact
effective threshold from the LD approximation is reached for
$Q^2\approx4$--$6$ GeV$^2$. As $Q^2$ increases~further,~the~exact
effective threshold approaches the LD limit quickly, thus
improving the accuracy of the LD model for the elastic form
factor. For generic confining interactions, the LD approach gives
very precise results for $Q^2\ge20$--$30$~GeV$^2$. In other words,
it is not possible to predict at which value $Q_*^2$ of the
momentum transfer the LD model provides~a good approximation
(with, say, 5\% accuracy) to the exact form factor, as the precise
$Q_*^2$ depends on subtle~details of the confining interactions.
However, one important conclusion may be drawn from our
quantum-mechanical analysis: {\it independently\/} of any subtle
details, the accuracy of the LD approximation {\it increases\/}
for $Q^2\ge4$--$6$~GeV$^2$.
\item 
Existing data on the pion elastic
form factor indicate that the LD limit $s_{\rm
LD}\equiv4\pi^2f_\pi^2$ of the effective threshold~may be reached
already at the relatively low values $Q^2\approx5$--$6$ GeV$^2$ of
the momentum transfer.
\end{enumerate}
On the basis of the results
obtained in this study, we are forced to conclude that those large
deviations from~the~LD~limit in the range $Q^2=20$--$50$~GeV$^2$
reported in \cite{recent1,recent2} appear to us highly unlikely.

Of course, our analysis does not provide a proof of but an
argument for the accuracy of the LD approximation~in QCD and it
gives some hint towards the behaviour of the pion elastic form
factor $F_\pi(Q^2)$ to be expected for large~values of $Q^2$.
Accordingly, the accurate experimental determination of
$F_\pi(Q^2)$ for $Q^2=4$--$10$ GeV$^2$ expected to be achieved by
JLab will have important implications for the predicted behaviour
of $F_\pi(Q^2)$ at large $Q^2$ up to asymptotically~high~momenta.

\acknowledgments 
We would like to thank Silvano Simula for
numerous interesting, enjoyable, and helpful discussions. D.~M.\
gratefully acknowledges support by the Austrian Science Fund (FWF)
under Project No.~P22843-N16. This work was supported in part by a
grant for leading scientific schools 1456.2008.2 as well as by
FASI State Contract No.~02.740.11.0244.

\appendix\section{\label{Sec:App}Perturbative expansion of Green functions in quantum mechanics}
We construct the perturbative
expansions of both polarization operator and vertex function in
quantum mechanics.

\subsection{Polarization operator}
\begin{figure}[h!]
\begin{center}
\includegraphics[width=12cm]{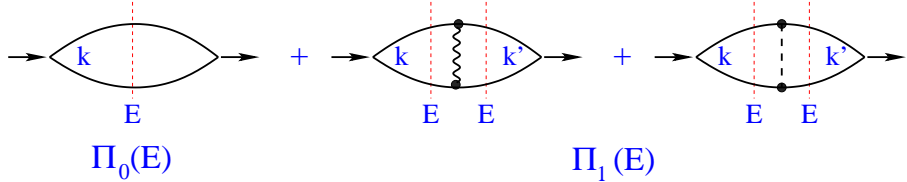}
\caption{\label{Fig:1}Expansion of the polarization operator in
terms of Coulomb (wavy line) and confining (dashed line)
interaction potentials.}
\end{center}
\end{figure}

The polarization operator
$\Pi(E)$ is defined by \cite{qmsr}
\begin{eqnarray}
\Pi(E)=\langle\bm{r}'=\bm{0}|G(E)|\bm{r}=\bm{0}\rangle,
\end{eqnarray}
where $G(E)$ is the full Green function, i.e., $G(E)=(H-E)^{-1},$
defined by the model Hamiltonian under consideration
\begin{eqnarray}
H=H_0+V(r),\qquad H_0\equiv\frac{\bm{k}^2}{2m},\qquad
r\equiv|\bm{r}|.
\end{eqnarray}
The expansion of the full Green
function $G(E)$ in powers of the interaction potential $V$ has the
well-known form
\begin{eqnarray}
\label{lippmann}
G(E)=G_0(E)-G_0(E)VG_0(E)+G_0(E)VG_0(E)VG_0(E)+\cdots,
\end{eqnarray}
with $G_0(E)=(H_0-E)^{-1}.$ It generates the corresponding
expansion of $\Pi(E)$:
\begin{eqnarray}
\label{seriesPi}
\Pi(E)=\Pi_0(E)+\Pi_1(E)+\cdots.\end{eqnarray}Explicitly, one
finds\begin{align}\Pi_0(E)&=\frac{1}{(2\pi)^3}\int\frac{{\rm
d}^3k}{\frac{\bm{k}^2}{2m}-E},\\
\Pi_1(E)&=-\frac{1}{(2\pi)^6}
\int\frac{{\rm d}^3k}{\frac{\bm{k}^2}{2m}-E}\frac{{\rm
d}^3k'}{\frac{\bm{k}'^2}{2m}-E}
V\!\left((\bm{k}-\bm{k}')^2\right).
\end{align}
We consider
interaction potentials $V(r)$ which consist of a Coulombic and a
confining part:
\begin{eqnarray}
V(r)=-\frac{\alpha}{r}+V_{\rm conf}(r).
\end{eqnarray}
Then the expansion (\ref{seriesPi}) becomes
a double expansion in powers of the Coulomb coupling $\alpha$ and
the confining potential $V_{\rm conf}$ (see Fig.~\ref{Fig:1}).

The contribution to $\Pi(E)$ arising from the Coulombic potential
is referred to as the perturbative contribution,~$\Pi_{\rm pert}$.
The contributions involving the confining potential $V_{\rm conf}$
(including the mixed terms receiving contributions from both
confining and Coulomb parts) are referred to as the power
corrections, $\Pi_{\rm power}.$ For instance, the first-order
perturbative contribution reads
\begin{eqnarray}
\Pi_1^{(\alpha)}(E)
=\frac{1}{(2\pi)^6}\int\frac{{\rm d}^3k}{\frac{\bm{k}^2}{2m}-E}
\frac{{\rm d}^3k'}{\frac{\bm{k}'^2}{2m}-E}
\frac{4\pi\alpha}{(\bm{k}-\bm{k}')^2}=\frac{\alpha m}{8\pi^2}
\int\frac{{\rm d}^3k}{\left(\frac{\bm{k}^2}{2m}-E\right)|\bm{k}|}.
\end{eqnarray}
The integral diverges but becomes convergent after applying the Borel
transformation $1/(a-E)\to\exp(-aT)$.\footnote{Note that the Borel
transform of the Green function $(H-E)^{-1}$ yields the
quantum-mechanical time-evolution operator in imaginary~time
$U(T)=\exp(-HT)$.} The Borel-transformed polarization operator
$\Pi(T)$ has the form
\begin{eqnarray}
\Pi(T)=\Pi_{\rm pert}(T)+\Pi_{\rm power}(T), \qquad\Pi_{\rm
pert}(T)=\left(\frac{m}{2\pi T}\right)^{3/2} \left[1+\sqrt{2\pi
mT}\alpha+\frac{1}{3}m\pi^2T\alpha^2
+O(\alpha^3)\right].
\end{eqnarray}
In the LD limit, that is, for $T\to0$, only $\Pi_{\rm pert}(T)$ will be relevant. 
Nevertheless,
as an illustration we provide also the result for the power
corrections $\Pi_{\rm power}(T)$ for the case of a
harmonic-oscillator confining potential $V_{\rm conf}(r)=m\omega^2r^2/2$:
\begin{eqnarray}
\Pi_{\rm power}(T)=\left(\frac{m}{2\pi T}\right)^{3/2}
\left[-\frac{1}{4}\omega^2T^2\left(1+\frac{11}{12}\sqrt{2\pi
mT}\alpha\right)+\frac{19}{480}\omega^4T^4\right].
\end{eqnarray}
Let us point out that $\Pi_{\rm power}(T=0)$ vanishes, similar to QCD.
The radiative corrections in $\Pi_{\rm pert}(T)$ have a
less~singular behaviour compared to the free Green function, so
the system behaves as quasi-free system. In QCD such a behaviour,
frequently regarded as an indication of asymptotic freedom, occurs
due to the running of the strong coupling $\alpha_s$~and~its
vanishing at small distances. Interestingly, in the
nonrelativistic potential model this feature is built-in
automatically.

Now, according to the standard procedures of the method of sum
rules, the dual correlator is obtained by applying~a low-energy
cut at some threshold $\bar k_{\rm eff}$ in the spectral
representation for the perturbative contribution to the
correlator:
\begin{eqnarray}
\Pi_{\rm dual}(T,\bar k_{\rm eff})=
\frac{1}{2\pi^2}\int\limits_0^{\bar k_{\rm eff}}{\rm d}k\,k^2
\exp\!\left(-\frac{k^2}{2m}T\right)\!\left[1+\frac{\pi m\alpha}{k}
+\frac{(\pi m\alpha)^2}{3k^2}+O(\alpha^3)\right]+\Pi_{\rm power}(T).
\end{eqnarray}
By construction, the dual correlator
$\Pi_{\rm dual}(T,\bar k_{\rm eff})$ is related to the
ground-state contribution by
\begin{eqnarray}
\label{Pidual}
\Pi_{\rm dual}(T,\bar k_{\rm eff})=\Pi_{\rm g}(T)\equiv R_{\rm
g}\exp(-E_{\rm g}T),\qquad R_{\rm g}\equiv|\psi_{\rm g}(r=0)|^2.
\end{eqnarray}
As we have shown in our previous studies
of potential models, the effective continuum threshold defined
according to (\ref{Pidual}) is a function of the Borel time
parameter $T$. For $T=0,$ one finds
\begin{eqnarray}
\label{picut}
\Pi_{\rm dual}(\bar k_{\rm eff},T=0)=\frac{1}{6\pi^2}\,{\bar k_{\rm eff}}^3+\frac{\alpha m}{4\pi}\,\bar k_{\rm eff}^2+\cdots.
\end{eqnarray}

\subsection{Vertex function}
\begin{figure}[h!]
\begin{center}
\includegraphics[width=12cm]{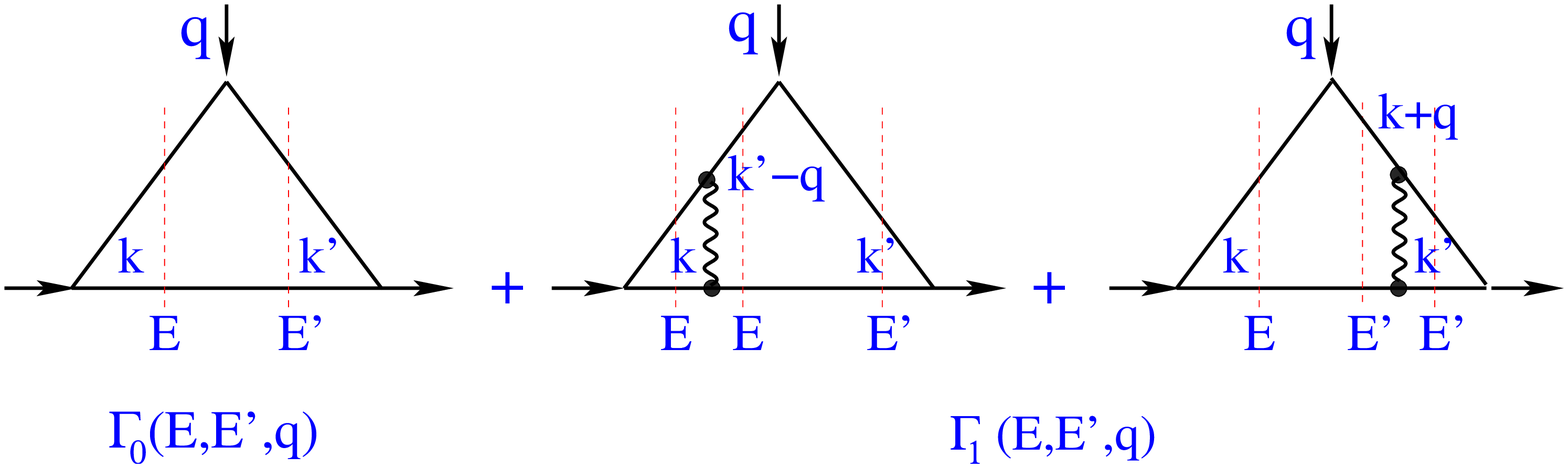}
\caption{\label{Fig:7}Nonrelativistic Feynman diagrams
representing the lowest perturbative contributions to the vertex
function $\Gamma(E,E',Q)$.}
\end{center}
\end{figure}

We now calculate the vertex function $\Gamma(E,E',Q),$ defined by
\begin{eqnarray}
\label{Gamma3pt}
\Gamma(E,E',Q)=\langle\bm{r}'=\bm{0}|G(E)J(\bm{q})G(E')|\bm{r}=\bm{0}\rangle,
\qquad Q\equiv|\bm{q}|,
\end{eqnarray}
where $J(\bm{q})$ is the
operator which adds a momentum $\bm{q}$ to the interacting
constituent. The expansions (\ref{lippmann}) of the~full Green
functions $G(E)$ and $G(E')$ in powers of the interaction entail a
corresponding expansion of
$\Gamma(E,E',Q),$~cf.~Fig.~\ref{Fig:7}:
\begin{eqnarray}
\label{Eq:vertexp}
\Gamma(E,E',Q)=\Gamma_0(E,E',Q)+\Gamma_1(E,E',Q)+\cdots.
\end{eqnarray}

For the vertex functions $\Gamma_i(E,E',Q),$ $i=0,1,\dots,$ in
(\ref{Eq:vertexp}), their double spectral representations may be
written~as
\begin{eqnarray}
\label{Gamma_double}
\Gamma_i(E,E',Q)=\int\frac{{\rm d}z}{\frac{z}{2m}-E}\frac{{\rm
d}z'}{\frac{z'}{2m}-E'}\,\Delta_i(z,z',Q).
\end{eqnarray}
The vertex functions $\Gamma_i(E,E',Q=0)$ and the polarization operators
$\Pi_i(E)$ satisfy the Ward identities
\begin{eqnarray}
\label{WI}
\Gamma_i(E,E',Q=0)=\frac{\Pi_i(E)-\Pi_i(E')}{E-E'},
\end{eqnarray}
which are equivalent to the following relations between the
corresponding spectral densities:
\begin{eqnarray}
\label{WI1}
\lim\limits_{Q\to0}\Delta_i(z,z',Q)=\delta(z-z')\,\rho_i(z).
\end{eqnarray}
We shall consider the double Borel transform $E\to T$ and $E'\to
T'$: $(a-E)^{-1}\to\exp(-aT),$ $(a'-E')^{-1}\to\exp(-a'T')$.
Equation (\ref{WI1}) leads to the following Ward identities for
the Borel images:
\begin{eqnarray}
\label{WI2}
\Gamma_i(T,T',Q=0)=\Pi_i(T+T').
\end{eqnarray}

\subsubsection{One-loop contribution $\Gamma_0$ to the vertex function}
The zero-order one-loop term has the form
\begin{eqnarray}
\Gamma_0(E,E',Q) =\frac{1}{(2\pi)^3}\int\frac{{\rm
d}^3k}{\left(\frac{\bm{ k}^2}{2m}-E\right)\!
\left(\frac{(\bm{k}+\bm{q})^2}{2m}-E'\right)}.
\end{eqnarray}
This may be written as the double spectral representation
\begin{eqnarray}
\Gamma_0(E,E',Q)=\int\frac{{\rm d}z}{\frac{z}{2m}-E}\frac{{\rm
d}z'}{\frac{z'}{2m}-E'}\,\Delta_0(z,z',Q),
\end{eqnarray}
where
\begin{eqnarray}
\Delta_0(z,z',Q)=\frac{1}{(2\pi)^3}\int{\rm d}^3k\,
\delta\!\left(z-\bm{k}^2\right)\delta\!\left(z'-(\bm{k}-\bm{q})^2\right)
=\frac{1}{(2\pi)^3}\frac{\pi}{2Q}\,\theta\!\left((z'-z-Q^2)^2-4zQ^2<0\right).
\end{eqnarray}
Hereafter, we use the notations $k\equiv\sqrt{z}$ and
$k'\equiv\sqrt{z'}$. In terms of the variables $k$ and $k',$ the
$\theta$ function takes the~form
\begin{eqnarray}
\theta\!\left((z'-z-Q^2)^2-4zQ^2<0\right)
=\theta\!\left(|k-Q|<k'<k+Q\right).
\end{eqnarray}

\subsubsection{Two-loop contribution $\Gamma_1$ to the vertex function}
We consider here only corrections related to the {\em Coulomb\/} potential, 
since power corrections induced by the
confining interaction vanish in the LD limit. The two-loop
$O(\alpha)$ correction receives two contributions and has the form
\begin{eqnarray}
\Gamma_1(E,E',Q)=
\frac{1}{(2\pi)^6}\int\frac{{\rm d}^3k}{\frac{\bm{k}^2}{2m}-E}
\frac{{\rm d}^3k'}{\frac{\bm{k}'^2}{2m}-E'}
\frac{4\pi\alpha}{\left(\bm{k}-(\bm{k}'-\bm{q})\right)^2}
\left[\frac{1}{\frac{(\bm{k}'-\bm{q})^2}{2m}-E'}
+\frac{1}{\frac{(\bm{k}+\bm{q})^2}{2m}-E}\right].
\end{eqnarray}
Having in mind the subsequent application of a double
Borel transformation in $E$ and $E',$ it is convenient to
represent $\Gamma_1$ as a sum of two terms,
$\Gamma_1=\Gamma_1^{(a)}+\Gamma_1^{(b)}$, with
\begin{align}
\Gamma_1^{(a)}(E,E',Q)&=\frac{\alpha m}{8\pi^5}\int\frac{{\rm
d}^3k'}{\left(\frac{\bm{k}'^2}{2m}-E'\right)
\left(\frac{(\bm{k}'-\bm{q})^2}{2m}-E\right)}\int\frac{{\rm
d}^3k}{\left(\bm{k}-(\bm{k}'-\bm{q})\right)^2
\left[\bm{k}^2-(\bm{k}'-\bm{q})^2\right]}\nonumber\\
&+\frac{\alpha m}{8\pi^5}\int\frac{{\rm
d}^3k}{\left(\frac{\bm{k}^2}{2m}-E\right)
\left(\frac{(\bm{k}+\bm{q})^2}{2m}-E'\right)}\int\frac{{\rm
d}^3k'}{\left((\bm{k}+\bm{q})-\bm{k}'\right)^2
\left[\bm{k}'^2-(\bm{k}+\bm{q})^2\right]},\nonumber\\
\label{A.25}
\Gamma_1^{(b)}(E,E',Q)&=\frac{\alpha m}{8\pi^5}\int\frac{{\rm
d}^3k}{\frac{\bm{k}^2}{2m}-E}\frac{{\rm
d}^3k'}{\frac{\bm{k}'^2}{2m}-E'}\frac{1}{(\bm{k}+\bm{q}-\bm{k}')^2}
\left[\frac{1}{(\bm{k}'-\bm{q})^2-\bm{k}^2}
+\frac{1}{(\bm{k}+\bm{q})^2-\bm{k}'^2}\right].
\end{align}
The double Borel transformation in $E\to T$ and $E'\to T'$ is now
easily performed.

Let us start with $\Gamma_1^{(a)}.$ One integration in
$\Gamma_1^{(a)}$ may be performed, leading to
\begin{eqnarray}
\Gamma_1^{(a)}(E,E',Q)=\frac{\alpha m}{16\pi^2}\left[\int
\frac{{\rm d}^3k'}{\left(\frac{(\bm{k}'+\bm{q})^2}{2m}-E'\right)\!
\left(\frac{\bm{k}'^2}{2m}-E\right)|\bm{k}'|}+\int\frac{{\rm
d}^3k}{\left(\frac{(\bm{k}-\bm{q})^2}{2m}-E\right)\!
\left(\frac{\bm{k}^2}{2m}-E'\right)|\bm{k}|}\right].
\end{eqnarray}
The first term corresponds to the contribution of the ``left''
two-loop diagram in Fig.~\ref{Fig:7}, i.e., with the
potential~before~the interaction with the current $J(\bm{q}),$
while the second term is represented by the ``right'' two-loop
diagram in Fig.~\ref{Fig:7}.~~The corresponding double spectral
densities have a form very similar to $\Delta_0$:
\begin{equation}
\Delta_{1L}^{(a)}(k,k',Q)=\frac{\alpha m}{16\pi}\frac{\pi}{2Q}
\frac{1}{k}\,\theta\!\left(|k-Q|<k'<k+Q\right)\theta(0<k)\,\theta(0<k'),
\quad\Delta_{1R}^{(a)}(k,k',Q)=\Delta_{1L}^{(a)}(k',k,Q).
\end{equation}

Explicit calculations yield the following double spectral
densities of the two contributions to $\Gamma_1^{(b)}$
related~to~the~``left'' and ``right'' two-loop diagrams in
Fig.~\ref{Fig:7}:
\begin{equation}
\Delta_{1L}^{(b)}(k,k',Q)=\frac{\alpha m}{32\pi^6}
\frac{1}{Qk}\left[\log^2\!\left(\left|\frac{k'-Q+k}{k'-Q-k}\right|\right)
-\log^2\!\left(\left|\frac{k'+Q+k}{k'+Q-k}\right|\right)\right],\quad
\Delta_{1R}^{(b)}(k,k',Q)=\Delta_{1L}^{(b)}(k',k,Q).
\end{equation}
At $Q=0$, $\Gamma_1^{(a)}(T,T',Q=0)$ satisfies the Ward identity,
$\Gamma_1^{(a)}(T,T',Q=0)=\Pi_1^{(\alpha)}(T+T')$, whereas
$\Gamma_1^{(b)}(T,T',Q=0)$ vanishes: $\Gamma_1^{(b)}(T,T',Q=0)=0.$
For large $Q$ and $T,T'\ne0$, $\Gamma_1^{(b)}(T,T',Q)$ assumes a
factorizable form (see Eq.~(\ref{A.25})):
\begin{eqnarray}
\Gamma_1^{(b)}(T,T',Q)\to\frac{16\pi\alpha
m}{Q^4}\,\Pi_0(T)\,\Pi_0(T').
\end{eqnarray}
At the same time, both
$\Gamma_0(T,T',Q)$ and $\Gamma_1^{(a)}(T,T',Q)$ are exponentially
suppressed for large $Q$ and $T,T'\ne0$. Hence, $\Gamma_1^{(b)}$
determines the large-$Q$ behaviour of the vertex function.

\subsubsection{Dual correlator}
The dual correlator $\Gamma_{\rm dual}(T,T',Q)$ is constructed in a standard way, 
by application of a low-energy cut to the~double spectral representation of the
perturbative contribution to (\ref{Gamma_double}):
\begin{eqnarray}
\label{GammaBorel_double}\Gamma_{\rm dual}(T,T',Q)=
\int\limits_0^{k_{\rm eff}(Q,T)}{\rm d}k\,2k
\exp\!\left(-\frac{k^2}{2m}\,T\right)\int\limits_0^{k_{\rm
eff}(Q,T')}{\rm d}k'\,2k'\exp\!\left(-\frac{k'^2}{2m}\,T'\right)
\Delta(z,z',Q)+\Gamma_{\rm power}(T,T',Q).\qquad
\end{eqnarray}
By construction, the dual correlator corresponds to the ground-state
contribution $\exp(-E_g T)\exp(-E_g T')\,R_g\,F_g(Q).$

In the LD limit $T=0$ and $T'=0,$ $\Gamma_{\rm power}(T,T',Q)$
vanishes and the ground-state form factor $F_g(Q)$ is
related~to~the low-energy part of the perturbative contribution
considered above:
\begin{eqnarray}
\label{LD3ptsr}
\int\limits_0^{k_{\rm eff}(Q)}{\rm d}k\,2k\int\limits_0^{k_{\rm
eff}(Q)}{\rm d}k'\,2k'\Delta(k,k',Q)=F_{g}(Q)\,R_{g},
\end{eqnarray}
with
$\Delta(k,k',Q)=\Delta_0(k,k',Q)+\Delta_{1L}^{(a)}(k,k',Q)
+\Delta_{1R}^{(a)}(k,k',Q)+\Delta_{1L}^{(b)}(k,k',Q)
+\Delta_{1R}^{(b)}(k,k',Q)+O(\alpha^2)$.

In order to provide the correct normalization $F_g(Q=0)=1$ of the
elastic form factor $F_g(Q),$ the effective thresholds should be
related to each other according to $k_{\rm eff}(Q=0)=\bar k_{\rm
eff}$; then the form factor is correctly normalized due to~the
Ward identity (\ref{WI1}) satisfied by the spectral densities.

\end{document}